\documentclass[aps, groupedaddress, preprint]{revtex4-2}
\usepackage{amsmath, amssymb}
\usepackage{graphicx}
\usepackage[colorlinks=true, linkcolor=blue, citecolor=blue]{hyperref}
\setcitestyle{authoryear,round} 

\begin{document}
\title{Nonlinear asymptotic bubble growth in single-mode spherical Rayleigh–Taylor instability}

\author{De-Hua Zhang}
\affiliation{State Key Laboratory of High Temperature Gas Dynamics, School of Engineering Science, University of Science and Technology of China, Hefei 230026, China}

\author{Shi-Heng Wang}
\affiliation{State Key Laboratory of High Temperature Gas Dynamics, School of Engineering Science, University of Science and Technology of China, Hefei 230026, China}

\author{Ke-Jian Qian}
\affiliation{State Key Laboratory of High Temperature Gas Dynamics, School of Engineering Science, University of Science and Technology of China, Hefei 230026, China}

\author{Zhu-Jun Li}
\affiliation{State Key Laboratory of High Temperature Gas Dynamics, School of Engineering Science, University of Science and Technology of China, Hefei 230026, China}

\author{Rui Yan}
\email{ruiyan@ustc.edu.cn}
\affiliation{State Key Laboratory of High Temperature Gas Dynamics, School of Engineering Science, University of Science and Technology of China, Hefei 230026, China}

\author{Hang Ding}
\affiliation{State Key Laboratory of High Temperature Gas Dynamics, School of Engineering Science, University of Science and Technology of China, Hefei 230026, China}
\date{\today}

\begin{abstract}
We present an analytical model for the nonlinear growth of a single-mode Rayleigh–Taylor instability (RTI) bubble in spherical geometry. The model captures the bubble growth along the polar axis, spanning the linear to nonlinear regimes, for arbitrary Atwood numbers and under both converging- and diverging-gravity configurations. The model predicts that the bubble acceleration approaches an asymptotic value in the nonlinear stage. The spherical geometry is found to enhance the RTI bubble growth relative to planar and cylindrical configurations with the same effective perturbation wavenumber in the converging-gravity cases, whereas it mitigates the bubble growth in the diverging-gravity cases. The model predictions show favorable agreement with direct numerical simulations.
\end{abstract}

\maketitle
\newpage
	
	
	\section{Introduction}\label{sec:intro}
	The Rayleigh-Taylor instability (RTI) \citep{rayleigh1900,taylor1950} is a fundamental hydrodynamic instability occurring at the interface between heavy and light fluids when the heavy fluid is supported by the light fluid against gravity. In the planar geometry, i. e., the interface is a plane and the gravity is perpendicular to the interface, a sinusoidal perturbation with a wavelength of $\lambda$ grows exponentially with time $t$ at a growth rate $\gamma = \sqrt{A_T k g}$ in the limit of small perturbation amplitudes $\eta$, where $k\equiv 2\pi /\lambda$ is the perturbation wave number, $g$ is the value of gravity, $A_T = (\rho_H - \rho_L)/(\rho_H + \rho_L)$ is the Atwood number, $\rho_H$ and $\rho_L$ are the densities of the heavy and light fluids, respectively \citep{chandrasekhar1961}. This exponentially growing phase where $\eta \propto \exp[\gamma t]$ in the small perturbation limit, i. e., $k \eta \ll 1$, is usually referred to as the linear phase of RTI. As the perturbation amplitudes grow larger, RTI gradually enters the nonlinear phase where the interface forms the bubbles of the light fluid rising into the heavy fluid and the spikes of the heavy fluid falling into the light fluid. RTI plays important roles in astrophysics \citep{inogamov1999} and inertial confinement fusion (ICF) \citep{zhouye2025}, and analytical modeling the RTI bubble evolution has been attracting great research interest.
	
	For RTI in the planar geometry, analytical models describing the nonlinear bubble evolution \citep{layzer1955,mikaelian1998,goncharov2002,mikaelian2003,sohn2003} have been readily established based on potential-flow theories. The theories predicted that the bubble velocity saturates to a constant terminal velocity at $U_b = \sqrt{(2 A_T g)/[C_g (1+A_T) k]}$ \citep{goncharov2002}, with the dimension factor $C_g = 3$ and 1 for the two- and three-dimensional (2D and 3D) planar geometries, respectively. The RTI bubble evolution in the 2D cylindrical geometry has been analytically modeled by \cite{zhao2020} that includes effects due to a convergence geometry. \cite{zhao2020} predicted that the RTI bubble in the cylindrical geometry does not have a constant terminal velocity, but instead approaches a constant terminal acceleration $a_b^{\rm cyl}$, which agrees well with simulations.
	
	RTI in ICF and many astrophysical scenarios generally develop on spherical interfaces, thus modeling RTI in the spherical geometry (spherical RTI) would be more relevant to these regimes. Specifically, in the novel Double-Cone-Ignition (DCI) scheme \citep{zhang2020dci} of ICF, a pair of conical targets each of which encompasses a portion of a sphere are employed and the RTI bubbles near the central axis pose a significant threat to the integrity of the targets. Spherical RTI has also recently drawn increasing interest in the field of Extreme Ultraviolet (EUV) lithography, and  RTI occurring on the axis of the tin droplet in the early stage has been proposed  to account for sheet destabilization later on\citep{klein2020,qian2026}. 
	Modeling the bubble evolution in the spherical geometry is therefore desired for better understanding and evaluating the influence of RTI in these regimes. Weakly-nonlinear theories work fairly well in describing single-mode evolution \citep{zhangjing2017}, multi-mode couplings \citep{zhangjing2018}, and Bell–Plesset effects \citep{lixilai2026}, within the transition regime from linear to nonlinear phases of spherical RTI. 
	However, weakly-nonlinear theories are still limited to small interface deformations and unable to describe the asymptotic behaviours that are important to long-term predictions. 
	
	In this paper, we investigate the nonlinear growth of a bubble evolving from the single-mode spherical RTI, with particular emphasis on its asymptotic behaviours. By expanding the flow quantities in the vicinity surrounding the bubble tip moving along the polar axis, we develop a potential-flow model applicable to arbitrary Atwood numbers. The model is then used to reveal the influence of spherical geometry on the nonlinear bubble evolution and to clarify its substantial differences from the planar or cylindrical RTI.

	\section{Analytical model}\label{sec:model}
	\subsection{Two scenarios}
	\begin{figure}
		\begin{center}
			\includegraphics[width=12.5cm]{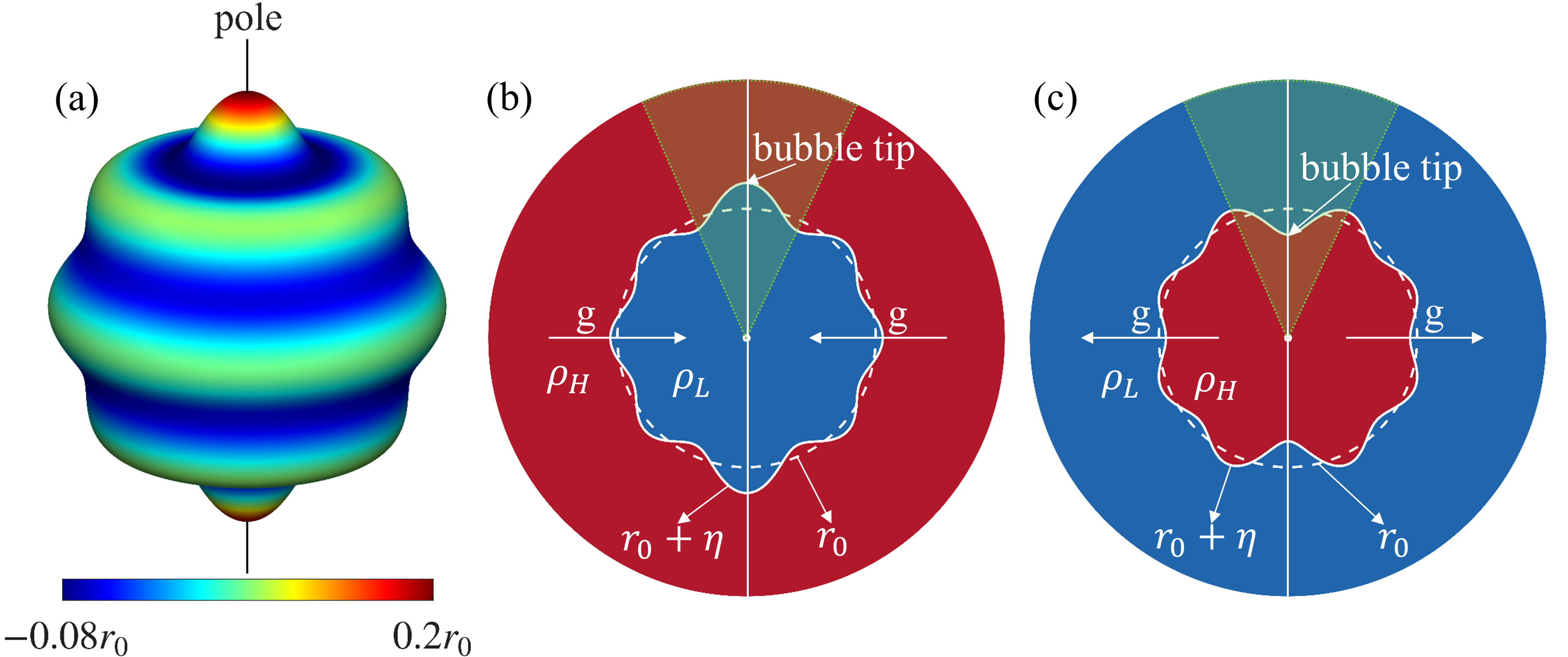}
			\caption{
				Schematic of a single-mode RTI in spherical geometry. 
				(a) Initial single-mode perturbation with the amplitude $\eta(\theta,0)=\eta_0(0)P_l(\cos\theta)$ on the spherical interface with $l=8$ and $\eta_0(0)=0.2 r_0$. The color map shows the radial displacement from the unperturbed spherical interface.
				(b) The converging-gravity (CG) configuration, also an axial cross-section of (a). 
				(c) The diverging-gravity (DG)  configuration.
				The heavy (with $\rho_H$, red) and light (with $\rho_L$, blue) fluids are divided by the perturbed interfaces (solid lines) in (b) and (c), the dashed lines are the unperturbed spherical interfaces at $r=r_0$, and the  masked conical areas are the simulation domains.
			}
			\label{fig:schematic}
		\end{center}
	\end{figure}
	
	We consider two irrotational, incompressible, and inviscid fluids in an axisymmetric spherical geometry, as illustrated in figure~\ref{fig:schematic}. The heavy and light fluids are denoted by the subscripts $H$ and $L$, respectively. 
	The unperturbed interface between the heavy and light fluids is located at $r=r_0$ (the dashed circles in figure~\ref{fig:schematic}). The axisymmetrically perturbed interface is described by $r=r_0+\eta(\theta,t)$, where $\eta(\theta,t)$ denotes the radial interface displacement as a function of the polar angle $\theta$ and time $t$. The initial perturbation is introduced as a single Legendre mode as shown in figure~\ref{fig:schematic}(a), $\eta(\theta,0)=\eta_0(0)P_l(\cos\theta)$, where $P_l(x)$ is the Legendre polynomial of degree $l$ [e. g., $P_0(x) = 1$, $P_1(x) = x$, $P_2(x) = (3 x^2 -1)/2$, ...], and $\eta_0(0)$ is the initial perturbation amplitude on the pole. 
	
	Two Rayleigh-Taylor unstable scenarios are explored depending on the orientation of the gravity ($\mathbf{g}$) applied, as illustrated by figure \ref{fig:schematic}(b) and (c). Figure \ref{fig:schematic}(b) demonstrates the converging-gravity (CG) scenario in which the gravity is pointing to the centre in a converging manner in the form of $\mathbf{g}=-g\mathbf{e}_r$, where $\mathbf{e}_r$ is the unit vector along the radial direction, $g$ is set to a positive constant. The heavier fluid is located in the outer side of the interface and the RTI bubble moves away from the centre in the CG configuration. Figure \ref{fig:schematic}(c) demonstrates the diverging-gravity (DG) scenario in which the gravity is directed outward in a diverging manner in the form of $\mathbf{g}= g\mathbf{e}_r$. The heavier fluid is located in the inner side of the interface and the RTI bubble moves towards the centre for DG. The linear growth rates for spherical RTI in both configurations have been well known \citep{zhangjing2017} as
	\begin{align}
		\label{eq:linear_rate}
		\gamma =\sqrt{ \frac{l(l+1)(\rho_{H}-\rho_{L})g}{[l(\rho_{H}+\rho_{L})+\rho_{in} ]r_0} },
	\end{align}
	where $\rho_{in}$ is the density of the fluid inside the interface, i. e., $\rho_{in} = \rho_{L}$ for CG and $\rho_{in} = \rho_H$ for DG. In \S~\ref{sec:sim}, the exponential growths in the linear phase will also be demonstrated to be recovered by our model derived in the following \S~\ref{sec:potential_model}. 
	
	\subsection{Potential-flow model}\label{sec:potential_model}
	Under the assumptions of inviscid, incompressible, irrotational flow,     
	\textcolor{blue}{and in the absence of surface tension}, 
	the velocities of the heavy and light fluids can be expressed in terms of velocity potentials as $\mathbf{u}_H=\nabla\phi_H$ and $\mathbf{u}_L=\nabla\phi_L$, where $\phi_H$ and $\phi_L$ are the velocity potentials of the heavy and light fluids, respectively. The velocity potentials satisfy the Laplace equations
	\begin{align}
		\label{eq:laplace}
		\nabla^2 \phi_H=\nabla^2 \phi_L=0.
	\end{align}
	At the axisymmetrically perturbed interface $r=r_0+\eta(\theta,t)$, $\phi_H$ and $\phi_L$ also have to satisfy the kinematic and dynamic boundary conditions\citep{Kundu2024}, which enforces the continuity of the velocity component normal to the interface and the pressure at the interface, respectively.
	The kinematic and dynamic boundary conditions at the perturbed interface can be expressed as
	\begin{align}
		\label{eq:BC1}
		\frac{\partial \eta}{\partial t}
		=
		\frac{\partial \phi_H}{\partial r}
		-\frac{1}{r^2}\frac{\partial \eta}{\partial \theta}\frac{\partial \phi_H}{\partial \theta}&,\\
		\label{eq:BC2}
		\frac{\partial \eta}{\partial t}
		=\frac{\partial \phi_L}{\partial r}-\frac{1}{r^2}
		\frac{\partial \eta}{\partial \theta}
		\frac{\partial \phi_L}{\partial \theta}&,\\
		\label{eq:BC3}
		\rho_{H}\left(\frac{\partial \phi_H}{\partial t}+\frac{|\nabla \phi_H|^2}{2} +\Psi \right)-\rho_{L}\left(\frac{\partial \phi_L}{\partial t}+\frac{|\nabla \phi_L|^2}{2}+\Psi \right)&=f(t),
	\end{align}
	where $f(t)$ is a function coming from the integration leading to the Bernoulli-like equation \eqref{eq:BC3}.
	The potential of the gravity field is given by $\Psi=gr$ for the  CG case and $\Psi=-gr$ for the DG case, respectively.
	
	Similar to the methodology of \cite{layzer1955} and \cite{goncharov2002}, we expand the interface amplitude $\eta$ near the bubble tip located at $\{r,\theta\}=\{r_0+\eta(0,t),0\}$ up to the second order of $\theta$, i.e., 
	$\eta(\theta,t) = \eta_0(t)+\eta_2(t)\theta^2$, where  $\eta_0(t)$ denotes the displacement of the bubble tip along the polar axis relative to the initial unperturbed interface ($r=r_0$),  $\eta_2(t)$ is associated with the local interfacial curvature on the bubble tip, and the term to the first order of $\theta$ naturally vanishes due to the symmetry with respect to the polar axis. We also expand \eqref{eq:BC1}-\eqref{eq:BC3} in the similar manner and obtain six differential equations grouping the zeroth- and second-order terms of $\theta$. Therefore, 
	in addition to $\eta_0(t)$, $\eta_2(t)$, and $f(t)$, three additional unknowns are allowed in the expressions of the velocity potentials to close the equation system. We then assume the velocity potentials near the bubble tip in the form of eigen functions satisfying the Laplace equations \eqref{eq:laplace}:
	\begin{align}
		\label{eq:phi_h_expression}
		\phi_H = a_1(t)\mathrm{P}_l(\cos\theta) \left(\frac{\eta_0}{r}\right)^{S_H},\\
		\label{eq:phi_l_expression}
		\phi_L = b_1(t)\mathrm{P}_l(\cos\theta) \left(\frac{r}{\eta_0}\right)^{S_L} + b_2(t)\frac{\eta_0}{r},
	\end{align}
	where $a_1$, $b_1$ and $b_2$ serve as the three unknowns to be determined by boundary conditions, the power indices $S_H=l+1$ and $S_L=-l$ are for CG, whereas $S_H=l$ and $S_L=-(l+1)$ for DG. 
	It is worthy to note that although $\phi_L$ in \eqref{eq:phi_l_expression} unphysically goes to infinity as $r \to 0$, we only need \eqref{eq:phi_l_expression} be accurate enough in a region  around the bubble tip, to capture the nonlinear bubble-tip dynamics governed by the local flow structures.
	Substituting \eqref{eq:phi_h_expression} and \eqref{eq:phi_l_expression} into \eqref{eq:BC1}-\eqref{eq:BC3} and expanding near the bubble tip yields the evolution equations for $\eta_0(t)$ and $\eta_2(t)$ as follows.
	
	For the CG configuration, the resulting equations are
	\begin{align}
		\label{eq:eta2_expr_gin}
		\dot\eta_2 &= -\frac{(l+1)\dot\eta_0}{4(r_0+\eta_0)}\left[8\eta_2+l(r_0 + \eta_0)\right],\\
		\label{eq:eta0_expr_gin}
		\frac{H_1}{8\left[8\eta_2-l(r_0+\eta_0)\right]}\ddot\eta _{0}
		&+\frac{H_2}{16\left[8\eta_2-l(r_0+\eta_0)\right]^2} \dot\eta_{0}^2 +  A_{T}g\eta_{2}=0,
	\end{align}
	where $H_1$ and $H_2$ are 
	\begin{subequations}
		\begin{align}
			H_1&=-l(2l+1-A_T)(r_0+\eta_{0})^2+8A_{T}l(r_0+\eta_0)\eta_2 + 64A_{T}{\eta_{2}}^2,  \\
			H_2&=l^2\left[A_T(10l^2+14l+3)-8l^2-10l-3\right](r_0+\eta_0)^2 \nonumber \\
			&+16l\left[A_T(4l^2+7l+4)-6l^2-11l-4\right](r_0+\eta_0)\eta_2+ 128l(l+2)A_{T}\eta_2^2.
		\end{align}
	\end{subequations}
	Similarly, for the DG configuration,  we have the the following relations:
	\begin{align}
		\label{eq:eta2_expr_gout}
		&\dot\eta_2 = \frac{l\dot\eta_0}{4(r_0+\eta_0)}  \left[ 8\eta_2 - (l+1)(r_0+\eta_0) \right],\\
		\label{eq:eta0_expr_gout}
		\frac{H_1}{8{\left[8\eta_{2}+(l+1)(r_{0}+\eta _{0)}\right]}} &\ddot{\eta}_0
		+\frac{H_2}{16{\left[8\eta_{2}+(l+1)(r_{0}+\eta _{0})\right]}^2}\dot{\eta}_0^2 + A_{T}g\eta_{2}=0,
	\end{align}
	where $H_1$ and $H_2$ are
	\begin{subequations}
		\begin{align}
			H_1  &= (l+1)(2l+1+A_T)(r_0+\eta_0)^2 + 8A_T(l+1)(r_0+\eta_0)\eta_2-64A_T\eta_2^2,  \\
			H_2 &=[8l^2+6l+1-A_T(10 l^2 +6 l -1)] (l+1)^2(r_0+\eta_0)^2 \nonumber \\
			&+ 16(l+1)[ A_T(4l^2+l+1)-6l^2-l+1](r_0+\eta_0)\eta_2-128 (l^2-1)A_T\eta_2^2.
		\end{align}
	\end{subequations}

	Through the derivations above, we have obtained a bubble-growth model for single-mode spherical RTI. The model consists of the coupled differential equations for $\eta_0$ and $\eta_2$, expressed by \eqref{eq:eta2_expr_gin}-\eqref{eq:eta0_expr_gin} for CG  and \eqref{eq:eta2_expr_gout}-\eqref{eq:eta0_expr_gout} for DG. 
	This model provides a unified description of the bubble evolution from  linear to nonlinear regimes, and will be verified against direct numerical simulations (DNS) in \S ~\ref{sec:sim} shortly after the discussions on the asymptotic solutions of these equations. 
	
	\subsection{Asymptotic behaviours}\label{sec:asym}
	\begin{figure}
		\begin{center}
			\includegraphics[width=7cm]{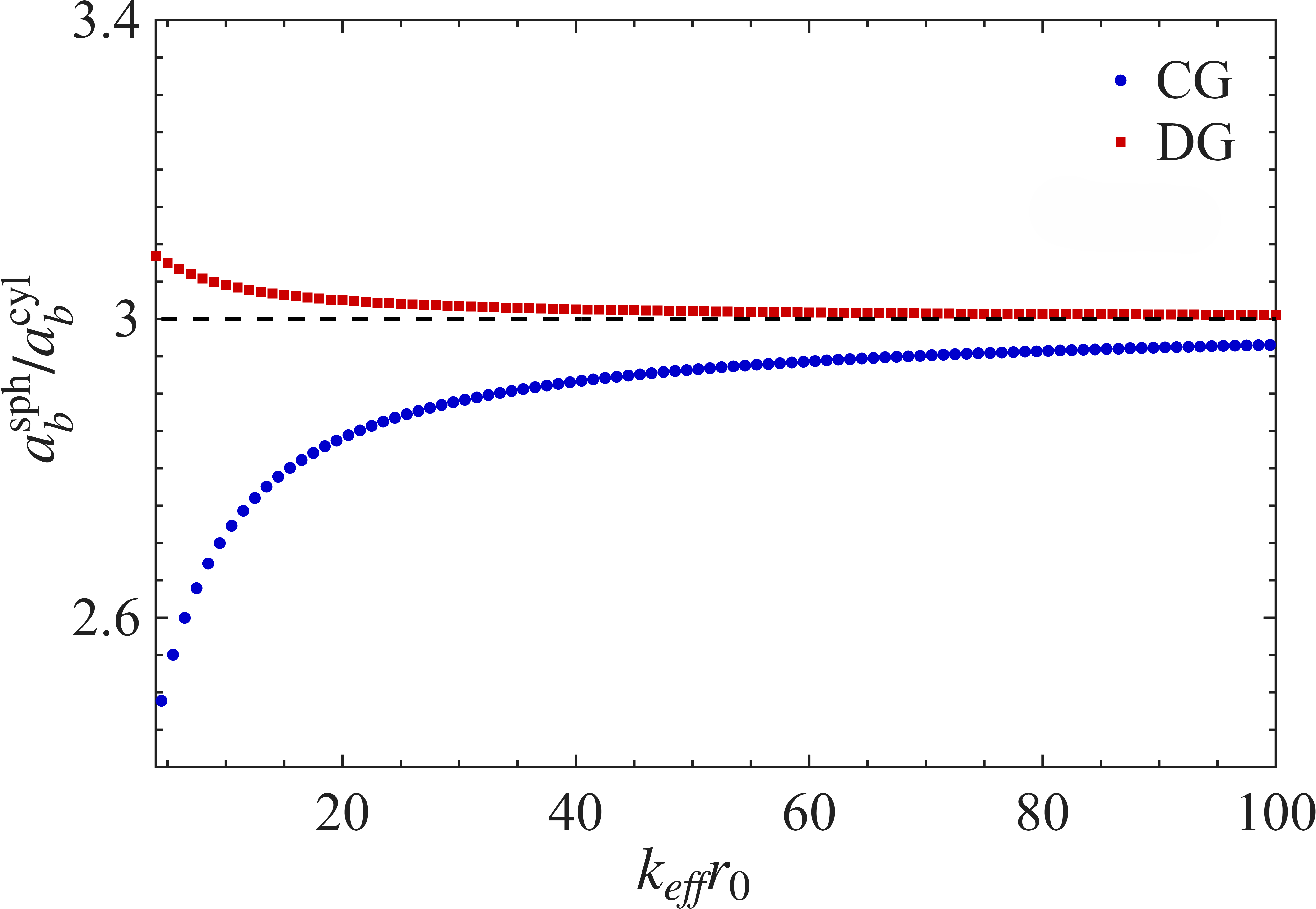}
			\caption{ Ratio of the asymptotic bubble accelerations with $A_T=0.3$, where $a_b^{\rm sph}=A/(B+2C)$ is from \eqref{eq:dUb},  $a_b^{\rm cyl}$ is the  bubble acceleration in the cylindrical geometry \citep{zhao2020}. 
				The dashed line shows the ratio equal to 3.
			}
			\label{fig:as_ac}
		\end{center}
	\end{figure}
	The influence of spherical geometry on the nonlinear bubble evolution can be revealed by the asymptotic behaviors of $\eta_0$ and $\eta_2$. We start from $\eta_2$, which characterizes the local interfacial curvature. The local expansion of the initial interface [$r = r_0 + \eta_0(0)P_l(\cos\theta)$] retaining the terms up to the second order of $\theta$ gives the relation between $\eta_2 (0)$ and $\eta_0 (0)$ as $\eta_2(0)=-l(l+1)\eta_0(0)/4$.
	Then, integrating \eqref{eq:eta2_expr_gin} and \eqref{eq:eta2_expr_gout} yields
	\begin{align}
		\label{eq:eta2_expr_ginout}
		\eta_2 = 
		\begin{cases}
			-\frac{l(l+1)}{4(2l+3)}\left[r_0+\eta_0-\left[r_0-2(l+1)\eta_0(0)\right]
			\left(\frac{r_0+\eta_0(0)}
			{r_0+\eta_0}\right)^{2l+2}\right], &\text{for CG}, \\
			\frac{l(l+1)}{4(2l-1)}\left[r_0+\eta_0-[r_0+2l\eta_0(0)]\left(\frac{r_0+\eta_0}{r_0+\eta_0(0)}\right)^{2l}\right],&\text{for DG}.
		\end{cases}
	\end{align}
	In the CG configuration, the bubble tip moves outward and $r_0+\eta_0(t)$ increases with time, whereas in the DG configuration the bubble tip moves inward and $r_0+\eta_0(t)$ decreases. 
	Consequently,
	in the limit of $t\rightarrow\infty$, the asymptotic relations $[r_0+\eta_0(0)]/[r_0+\eta_0(t)]\rightarrow0$ for CG and $\rightarrow \infty$ for DG should be satisfied. 
	Substituting the asymptotic relations into \eqref{eq:eta2_expr_ginout} yields the
	asymptotic limits for $\eta_2$ as
	\begin{align}
		\label{eq:eta2_t8}
		\eta_2 &\rightarrow
		\begin{cases} 
			\displaystyle -\frac{l(l+1)}{4(2l+3)}(r_0+\eta_0),
			& \text{for CG}, \\
			\displaystyle \frac{l(l+1)}{4(2l-1)}(r_0+\eta_0),
			& \text{for DG}.
		\end{cases}
	\end{align}
	Substituting \eqref{eq:eta2_t8} back into \eqref{eq:eta0_expr_gin} and
	\eqref{eq:eta0_expr_gout} yields an asymptotic evolution of $\eta_0$:
	\begin{align}
		\label{eq:eta0_expr_simplied}
		(r_0+\eta_0)A=(r_0+\eta_0)\ddot\eta_0 B + \dot \eta _{0}^2 C.
	\end{align} 
	For CG, the coefficients $A$, $B$ and $C$ are 
	\begin{subequations}
		\begin{align}
			\label{eq:eta0_A_gin}
			A &= \frac{l(l+1)A_{T}g}{4(2l+3)},\\
			\label{eq:eta0_B_gin}
			B &= \frac{(2l+1)(2l+3)^2-A_{T}(2l^2+10l+9)}{8(2l+3)(4\,l+5)},\\
			\label{eq:eta0_C_gin}
			C &=\frac{16l^4+72l^3+120l^2+86l+21 + A_{T}(16l^4+72l^3+90l^2+14l-21)}{{16(4l+5)}^2},
		\end{align} 
	\end{subequations}
	whereas for DG, the coefficients $A$, $B$ and $C$ are
	\begin{subequations}
		\begin{align}
			\label{eq:eta0_A_gout}
			A &= \frac{l(l+1)A_{T}g}{4(2l-1)},\\
			\label{eq:eta0_B_gout}
			B &= -\frac{(2l-1)^2(2l+1) +A_T(2l^2-6l+1)}{8(2l-1)(4l-1)},\\
			\label{eq:eta0_C_gout}
			C &=\frac{16l^4-8l^3+2l-1+A_T(16l^4-8l^3-30l^2 + 14l -1)}{16(4l-1)^2}.
		\end{align} 
	\end{subequations}
	Integrating \eqref{eq:eta0_expr_simplied}  yields
	\begin{align}
		\label{eq:eta0_final}
		\dot\eta_0^2=\frac{2A}{B+2C}\left(r_0+\eta_0(t)\right)
		+\left[\dot \eta_0(0)^2-\frac{2A}{B+2C}(r_0+\eta_0(0))\right]
		\left(\frac{r_0+\eta_0(0)}{r_0+\eta_0(t)}\right)^{\frac{2C}{B}}.
	\end{align}
	Taking the limit of $t\rightarrow\infty$,  
	together with the asymptotic
	conditions for $[r_0+\eta_0(0)]/[r_0+\eta_0(t)]$,
	we obtain the following asymptotic relation for $\eta_0$:
	\begin{align}
		\label{eq:Ub_gin}
		\frac{\dot \eta_0}{\sqrt{r_0+\eta_0}}\to\alpha\sqrt{\frac{2A}{B+2C}},
	\end{align}
	where $\alpha$ are 1 and -1 for the CG and DG configurations, respectively. Substituting \eqref{eq:Ub_gin} into \eqref{eq:eta0_expr_simplied} further yields the asymptotic value of the bubble acceleration ($\ddot \eta_0 $):
	\begin{align}
		\label{eq:dUb}
		\ddot \eta_0 \to \frac{A}{B+2C}.
	\end{align}
	
	Equations \eqref{eq:Ub_gin}-\eqref{eq:dUb} demonstrate that the nonlinear RTI bubble growth in the spherical geometry differs qualitatively from that in the planar geometry where the bubble velocity $\dot{\eta}_0$ asymptotically approaches a constant $U_b$ due to the balance between the buoyancy and gravity \citep{goncharov2002}. In contrast, the spherical geometry introduces an intrinsic time-dependent length scale associated with the instantaneous position of the bubble tip ($r_0+\eta_0$). It is $\dot{\eta}_0/\sqrt{r_0+\eta_0}$ rather than the bubble velocity $\dot{\eta}_0$ who asymptotically approaches a constant as $t\rightarrow\infty$, as shown by \eqref{eq:Ub_gin}. 
	Consistently, \eqref{eq:dUb}  predicts that the bubble acceleration approaches a constant $a_b^{\rm sph}=A/(B+2C) $, indicating that the nonlinear bubble evolution eventually enters a uniform-acceleration regime. It is worthy noting that $a_b^{\rm sph}$ is always positive, which means that a bubble is accelerating when it is moving outward in the CG case and decelerating when it is moving inward in the DG case. 
	This asymptotic-acceleration behavior is qualitatively similar to that in the cylindrical geometry where the bubble acceleration approaches a constant $a_b^{\rm cyl}$ \citep{zhao2020}.  
	
	Quantitative comparison between $a_b^{\rm sph}$ and $a_b^{\rm cyl}$ is demonstrated in figure~\ref{fig:as_ac}. For fair comparison on RTI in different geometries, an effective wavenumber $k_{\rm eff}$ is often defined \citep{guohongyu2018} to be  $k_{\rm eff} \equiv \sqrt{l(l+1)}/r_0$ for the spherical modes and $k_{\rm eff} \equiv n/r_0$ for the cylindrical modes, where $n$ is the number of perturbation wavelengths incorporated by the cylindrical circumference. For the same $k_{\rm eff}$, figure~\ref{fig:as_ac} shows that $a_b^{\rm sph}$ is significantly larger than $a_b^{\rm cyl}$ and the ratio $a_b^{\rm sph}/a_b^{\rm cyl}$ gradually approaches 3 for large $k_{\rm eff}$. The ratio of 3 is further found independent of $A_T$. Therefore, a spherical-RTI bubble experiences stronger asymptotic acceleration than its counterpart in the cylindrical geometry in the CG case, while its deceleration is also stronger than its cylindrical counterpart in the DG case.

	Furthermore, our model recovers the model by \cite{layzer1955} on an axisymmetric bubble growing on a planar interface and inside a  cylindrical tube, in the large-$l$ limit.
	To be analogous to the cylindrical tube, the RTI bubble in our model can be considered to grow within a cone $\theta\leq\Theta_l$, where $\Theta_l$ is the first minimum of the Legendre polynomial $P_l(\cos\theta)$ such that similar reflection boundaries apply on the walls of the cone and the tube. As $l$ increases, $\Theta_l$ decreases, making this conical region approach a cylindrical tube. As an asymptotic 
	property of the Legendre polynomials, $\Theta_l\to \beta_{1,1}/(l+1/2)$ as $l \to \infty$,
	where $\beta_{1,1} \approx 3.832$ is the first zero point of the first-order Bessel function. 
	Substituting this asymptotic relation into \eqref{eq:Ub_gin}  yields $|\dot{\eta}_0| \approx \sqrt{D A_T g/[(1+A_T)\beta_{1,1}]}$,
	where $D=2r_0\Theta_l$ is the equivalent diameter.
	For $A_T=1$, an asymptotic bubble velocity is obtained as $U_b \approx \sqrt{Dg/(2\beta_{1,1})}$,
	which recovers the prediction of \cite{layzer1955} for the saturated bubble velocity.

	\section{Numerical Verification}\label{sec:sim}
	\begin{figure}
		\begin{center}
			\includegraphics[width=9cm]{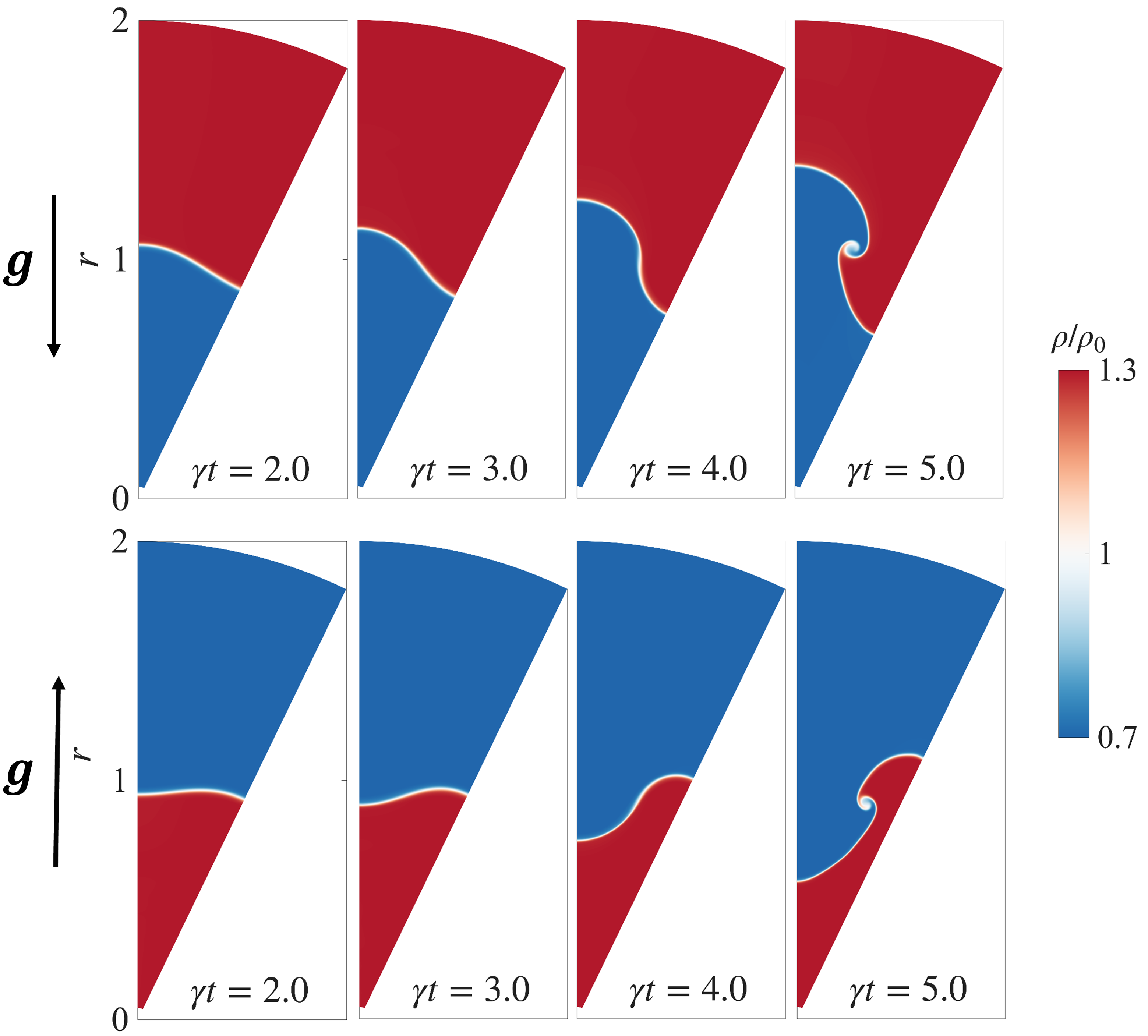}
			\caption{Density contours from the simulations with $A_T=0.3 $ and $ l=8$ at different times in the CG case [upper sub-figures], and in the DG case [lower sub-figures]. The linear growth rates $\gamma$ are calculated via \eqref{eq:linear_rate}.}
			\label{fig:dens_contours}
		\end{center}
	\end{figure}
	
	Numerical simulations of the spherical RTI are performed to validate the theoretical model developed in \S~\ref{sec:model}. The inviscid Euler equations are numerically solved:
	\begin{align}
		\label{eq:hydro_density}
		\frac{\partial \rho}{\partial t} +  \nabla \cdot( \rho \mathbf{u}) = 0,\\
		\label{eq:hydro_momentum}
		\frac{\partial \rho \mathbf{u}}{\partial t} +  \nabla \cdot(\rho \mathbf{u}\mathbf{u}) = \rho\mathbf{g},\\
		\label{eq:hydro_energy}
		\frac{\partial E}{\partial t} + \nabla \cdot(( E + p)\mathbf{u}) = \rho \mathbf{u}\cdot\mathbf{g},
	\end{align}
	where, $t$, $\rho$, $\mathbf{u}$, $p$, and $E$ denote the time, density, velocity, pressure, and total energy density, respectively. An ideal gas equation of state is adopted, with the total energy density given by $E=p/(\Gamma-1)+\rho |\mathbf{u}|^2/2$, where $\Gamma=1.4$ is the specific heat ratio. 
	The governing equations \eqref{eq:hydro_density}-\eqref{eq:hydro_energy} are solved using a finite-volume method. The spatial discretization is implemented with a fifth-order weighted essentially non-oscillatory (WENO) scheme~\citep{Jiang1996}, and the time integration is performed using a third-order explicit Runge-Kutta method. The HLLC approximate Riemann solver~\citep{Toro1994} is used to evaluate the numerical fluxes across cell boundaries.
	
	\begin{figure}
		\begin{center}
			\includegraphics[width=12cm]{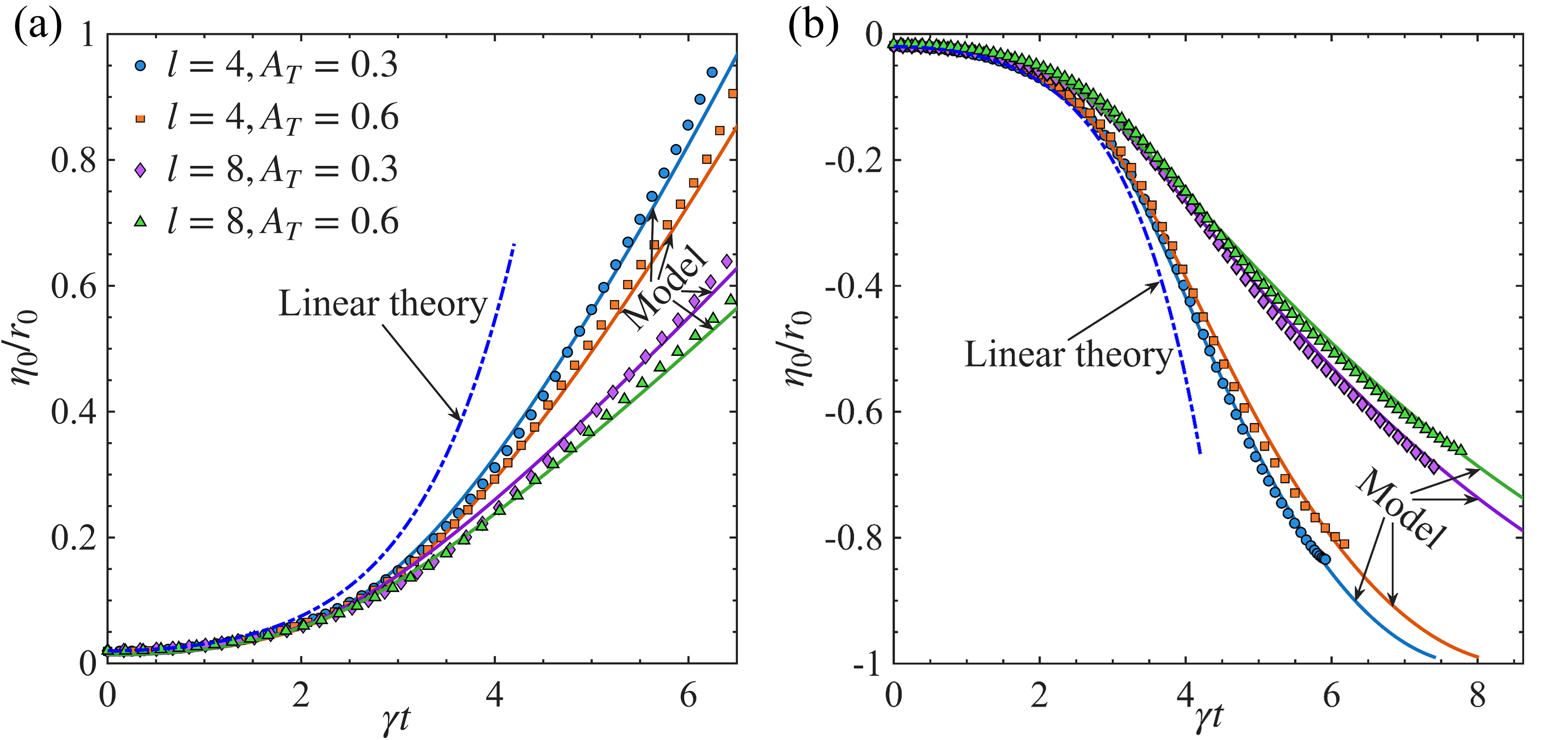}
			\caption{ Radial trajectory of the bubble tip in the CG ($a$) and DG ($b$) simulations. The solid lines and symbols are the results from the simulations and the present model, respectively. The linear theory predicting the growth of $\eta_0(t) = \eta_0(0)\cosh(\gamma t)$ is plotted in dashed lines.}
			\label{fig:eta0_model}
		\end{center}
	\end{figure}

	Axisymmetric simulations are performed in a spherical coordinate ($r,\theta$) for two Atwood numbers, $A_T=0.3$ and $0.6$, and two perturbation modes, $l=4$ and $8$.
	The perturbed interface dividing the heavy and light fluids is initialized to be a single Legendre mode $r=r_0+\eta_0(0)P_l(\cos\theta)$, with the amplitudes
	set to $\eta_0(0)=0.02r_0$ and $\eta_0(0)=-0.02r_0$ for the CG and DG cases, respectively. The fluids are initially static thus $\dot{\eta}_0(0) = 0$. An important dimensionless number $Fr\equiv u_0^2/(r_0 g)$ known as the Froude number is set to 10, where $u_0 = \sqrt{p_0/\rho_0}$ is a characteristic velocity calculated using the pressure at the interface ($p_0$) and $\rho_0 = (\rho_H + \rho_L)/2$.
	The simulation domain is set to be a conical region within  $\theta \le \Theta_l$ and $0.05 r_0 \le r \le 2.5 r_0$.
	The domain is discretized using $350\times1600$ and $184\times1600$ grid points for the $l=4$ and
	$l=8$ cases, respectively.
	To ensure the accuracy of the simulations, we have conducted a mesh convergence study using three sets of grids.
		
	Figure~\ref{fig:dens_contours} illustrates the evolution of the interface in a pair of CG and DG cases with $A_T=0.3$ and $l=8$. The upper sub-figures demonstrate an RTI bubble moving outwards in the CG case and the lower sub-figures demonstrate a bubble moving inwards in the DG case. 
	At $\gamma t=4$, nonlinear bubble structures have been well developed in both cases. 
	
	The predictions of the analytical model are compared with the simulation results for all considered combinations of $A_T$ and $l$. Figure~\ref{fig:eta0_model} compares the bubble-tip position $\eta_0$ obtained from the simulations with that predicted by numerically integrating \eqref{eq:eta2_expr_gin}-\eqref{eq:eta0_expr_gin} and \eqref{eq:eta2_expr_gout}-\eqref{eq:eta0_expr_gout} for the CG and DG configurations, respectively. At the early phase (e.g. $\gamma t<1.5$), the bubble growth agrees very well with the linear growth  [$\eta_0=\eta_0(0)\cosh(\gamma t)$],  
	and gradually deviates from the exponential growth as nonlinear effects become significant. Our model agrees well with the simulations throughout the linear and nonlinear stages with different  $A_T$ and $l$, evidencing the validity of the model.
	
\begin{figure}
		\begin{center}
			\includegraphics[width=12cm]{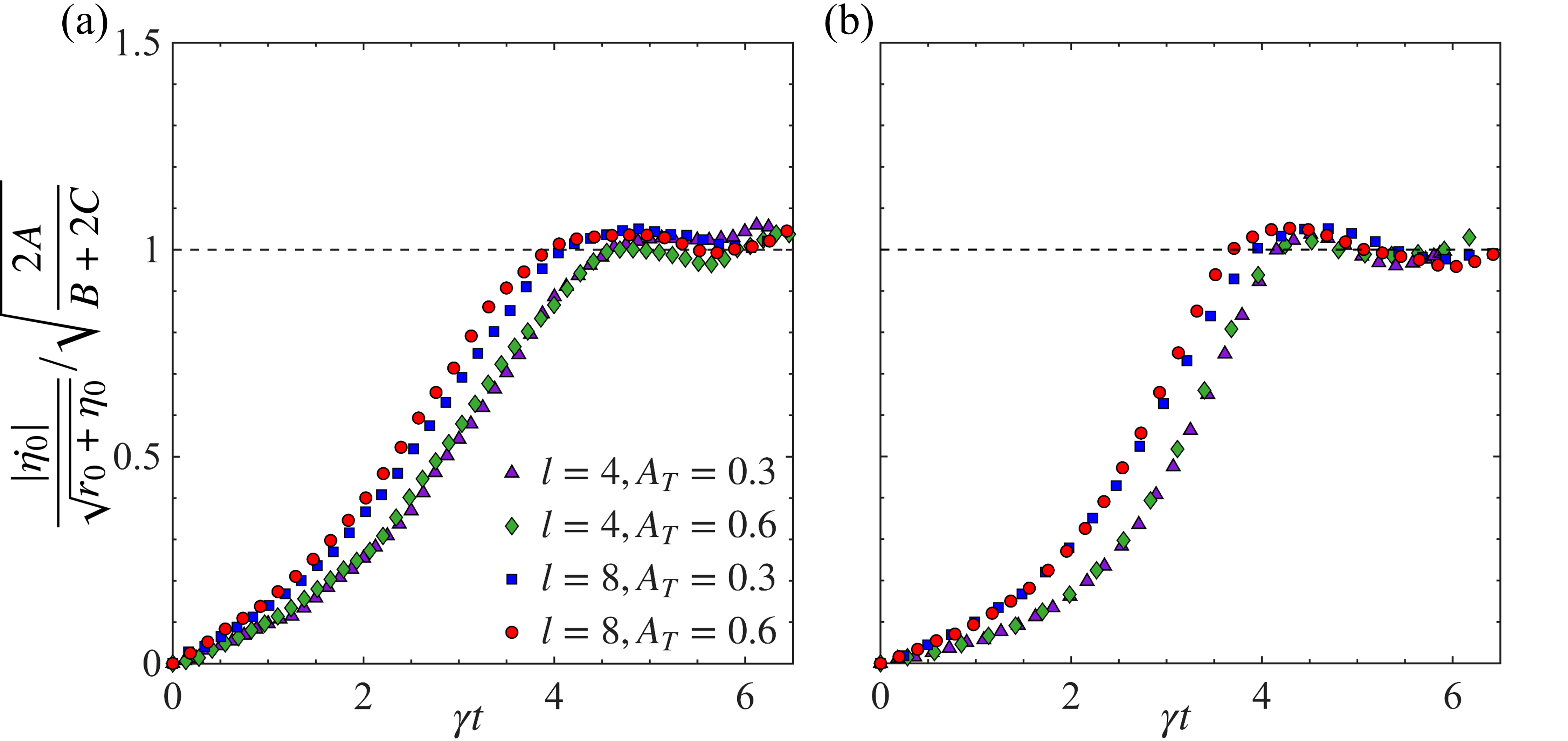}
			\caption{ 
				Evolution of $|\dot{\eta}_0|/\sqrt{r_0+\eta_0}$ for the CG ($a$) and DG ($b$) cases. The symbols are from the simulations and the dashed line is the asymptotic value predicted by the model via \eqref{eq:Ub_gin}. }
			\label{fig:Ub_model}
		\end{center}
	\end{figure}

	The analytical model predicts that the bubble growth in spherical RTI approaches an asymptotic regime characterized by a constant value of $|\dot{\eta}_0|/\sqrt{r_0+\eta_0}$ approaching
	$\sqrt{A/(B+2C)}$ [see \eqref{eq:Ub_gin} ]. Figure~\ref{fig:Ub_model} demonstrates such asymptotic behaviors in the simulations under both CG (a) and DG (b) configurations. The asymptotic regimes predicted by the theory are reached after several $e$-folding times ($\gamma t \approx 4$) in the simulations. 
	
	\section{Summary and conclusion}\label{sec:conclusion}
	In summary, we have developed an analytical bubble-growth model for spherical RTI. The model consists of coupled ordinary differential equations for the bubble-tip displacement $\eta_0$ and the interface curvature $\eta_2$ for both the CG and DG configurations, handles arbitrary $A_T \leq 1$, and provides a unified description of the bubble evolution from linear throughout to nonlinear regimes, with the effect of spherical geometry incorporated. 
	
	The asymptotic analysis of bubble evolution shows that both
	$\dot{\eta}_0/\sqrt{r_0+\eta_0}$ and $\ddot{\eta}_0$ approach constant values in the nonlinear regime. Consequently, the bubble accelerates while moving outward in the CG configuration and decelerates while moving inward in the DG configuration. For the same effective perturbation wavenumber $k_{\rm eff}$, the asymptotic acceleration in spherical geometry is larger than that in cylindrical geometry, with $a_b^{\rm sph}/a_b^{\rm cyl}\rightarrow3$ in the large $k_{\rm eff}$ limit, independently of $A_T$. Therefore, the model predicts that the spherical geometry substantially enhances the bubble growth in the CG configuration but mitigates it in the DG configuration, comparing to planar and cylindrical geometries. 
	The predictions have been verified against DNS and shown good agreement with the simulations.

	\bibliography{reference}   
	\bibliographystyle{plainnat}
\end{document}